%% file: sample-sigconf-authordraft.tex
\documentclass[sigconf]{acmart}

\AtBeginDocument{%
  }
\usepackage{multirow}
\setcopyright{acmlicensed}
\copyrightyear{2026}
\acmYear{2026}
\acmDOI{XXXXXXX.XXXXXXX}
\acmConference[NTCIR-19]{}{December 8--10, 2026}{Tokyo, Japan}
\acmISBN{978-1-4503-XXXX-X/2018/06}

\begin{document}

\title{Overview of the NTCIR-19 Automatic Evaluation of LLMs 2 (AEOLLM-2) Task}




\author{Junjie Chen}
\affiliation{%
  \institution{DCST, Tsinghua University}
  \department{Quan Cheng Laboratory}
  \city{Beijing 100084}
  \country{China}
}
\email{chenjj826@gmail.com}

\author{Yuxi Dong}
\affiliation{%
  \institution{University of Science and Technology Beijing}
  \city{Beijing 100084}
  \country{China}
}
\email{yuxidong2@gmail.com}

\author{Haitao Li}
\affiliation{%
  \institution{DCST, Tsinghua University}
  \department{Quan Cheng Laboratory}
  \city{Beijing 100084}
  \country{China}
}
\email{liht22@mails.tsinghua.edu.cn}

\author{Yiqun Liu}
\affiliation{%
  \institution{DCST, Tsinghua University}
  \department{Quan Cheng Laboratory}
  \city{Beijing 100084}
  \country{China}
}
\email{yiqunliu@tsinghua.edu.cn}

\author{Qingyao Ai}
\affiliation{%
  \institution{DCST, Tsinghua University}
  \department{Quan Cheng Laboratory}
  \city{Beijing 100084}
  \country{China}
}
\email{aiqy@tsinghua.edu.cn}


\begin{abstract}
In this paper, we provide an overview of the NTCIR-19 Automatic Evaluation of LLMs 2 (AEOLLM-2) task. Building on the success of the NTCIR-18 core task AEOLLM, we proposed AEOLLM-2 for NTCIR-19 to further investigate automatic evaluation methods for Large Language Models (LLMs), particularly in long-form text generation scenarios. In AEOLLM-2, we introduced a new subtask, Deep Research Evaluation, which focuses on the automatic evaluation of long-form deep research reports generated by LLMs. Participants developed evaluation methods to automatically assess the quality of these reports, and the performance of each method was measured by comparing its scores against human-annotated ground-truth labels. This year, we received 91 runs from 10 teams in total. This paper describes the background of the task, the dataset construction, the evaluation measures, the participants' methods, and the final evaluation results.
\end{abstract}




\maketitle

\section{Introduction}
\label{section one}
\input{intro}

\section{Task Framework}
\input{frame}
\section{Implementation Details}
\input{details}

\begin{table*}[t]

\caption{The results from the formal run on the reserved set. The best result is highlighted in bold, and the second-best result is underlined.}

\label{result}

\resizebox{\textwidth}{!}{%
\begin{tabular}{lccccccccccccccc}

\hline

\multicolumn{1}{c}{\multirow{2}{*}{TEAM}}
& \multicolumn{3}{c}{weighted overall}
& \multicolumn{3}{c}{comprehensiveness}
& \multicolumn{3}{c}{insight}
& \multicolumn{3}{c}{instruction\_following}
& \multicolumn{3}{c}{readability}
\\ \cline{2-16}

\multicolumn{1}{c}{}
& spearman & kendall & accuracy
& spearman & kendall & accuracy
& spearman & kendall & accuracy
& spearman & kendall & accuracy
& spearman & kendall & accuracy
\\ \hline

RSLIR
& \textbf{0.7071} & \textbf{0.5592} & \textbf{0.7796}
& \textbf{0.6881} & \textbf{0.5355} & \textbf{0.7533}
& \textbf{0.6890} & \textbf{0.5488} & \textbf{0.7425}
& \textbf{0.5353} & \textbf{0.4199} & \textbf{0.6350}
& 0.3978 & 0.2774 & 0.6342
\\

ISLab
& \underline{0.7024} & \underline{0.5555} & \underline{0.7775}
& \underline{0.6264} & \underline{0.4937} & 0.7163
& 0.6037 & 0.4822 & 0.6692
& \underline{0.5193} & \underline{0.4010} & \underline{0.6167}
& 0.3268 & 0.2515 & 0.5100
\\

TMUHLLA
& 0.6616 & 0.5242 & 0.7621
& 0.5724 & 0.4302 & 0.7063
& \underline{0.6537} & 0.5074 & \underline{0.7300}
& 0.4061 & 0.3054 & 0.6054
& 0.3609 & 0.2678 & 0.6308
\\

LINKLab
& 0.6554 & 0.5067 & 0.7533
& 0.5775 & 0.4397 & 0.6933
& 0.6438 & 0.4977 & 0.7104
& 0.4107 & 0.3227 & 0.5963
& 0.2814 & 0.1938 & 0.5733
\\

IITBHU
& 0.6080 & 0.4787 & 0.7375
& 0.5710 & 0.4459 & 0.6867
& 0.6434 & \underline{0.5143} & 0.7117
& 0.4123 & 0.3250 & 0.5725
& 0.4595 & \underline{0.3612} & 0.5542
\\

SCUNLP
& 0.5838 & 0.4450 & 0.7225
& 0.5133 & 0.3862 & 0.6842
& 0.5865 & 0.4399 & 0.6971
& 0.1993 & 0.1463 & 0.5250
& \textbf{0.5197} & \textbf{0.3827} & \textbf{0.6883}
\\

THUFAITH
& 0.5782 & 0.4375 & 0.7188
& 0.6252 & 0.4812 & \underline{0.7171}
& 0.3556 & 0.2764 & 0.6158
& 0.4258 & 0.3230 & 0.6063
& 0.3444 & 0.2536 & 0.5808
\\

kucs17
& 0.3319 & 0.2383 & 0.6192
& 0.1584 & 0.1067 & 0.5304
& 0.4352 & 0.3182 & 0.6300
& 0.1193 & 0.0745 & 0.3258
& 0.3971 & 0.2768 & \underline{0.6354}
\\ \hline

\multicolumn{16}{c}{Baseline Methods}
\\ \hline

glm-4.7
& 0.6177 & 0.4830 & 0.7413
& 0.5702 & 0.4407 & 0.6867
& 0.6312 & 0.5045 & 0.7175
& 0.4592 & 0.3604 & 0.5592
& \underline{0.4710} & 0.3610 & 0.6167
\\

deepseek-v3.2
& 0.6101 & 0.4739 & 0.7321
& 0.5781 & 0.4552 & 0.6796
& 0.6111 & 0.4903 & 0.6988
& 0.3484 & 0.2633 & 0.5354
& 0.4008 & 0.3217 & 0.4421
\\

qwen-max
& 0.5617 & 0.4342 & 0.6979
& 0.4646 & 0.3620 & 0.5454
& 0.5706 & 0.4663 & 0.5933
& 0.3202 & 0.2478 & 0.4883
& 0.3449 & 0.2777 & 0.4642
\\

gpt-4o-mini
& 0.5564 & 0.4128 & 0.6917
& 0.5494 & 0.4382 & 0.6154
& 0.5785 & 0.4654 & 0.6050
& 0.3647 & 0.2875 & 0.4538
& 0.2632 & 0.2096 & 0.3721
\\

gpt-4o-2024-11-20
& 0.5312 & 0.4032 & 0.6821
& 0.4466 & 0.3535 & 0.5746
& 0.5599 & 0.4386 & 0.6300
& 0.2801 & 0.2151 & 0.4917
& 0.3733 & 0.2919 & 0.5004
\\

qwen3-max
& 0.5299 & 0.4169 & 0.6692
& 0.4972 & 0.3956 & 0.5250
& 0.5762 & 0.4661 & 0.5871
& 0.3614 & 0.2916 & 0.4863
& 0.3808 & 0.3053 & 0.4067
\\ \hline

\end{tabular}%
}

\end{table*}
\section{Participants' Methods}
\input{method}

\section{Official Results}
\input{result}

\section{Conclusion}
This paper provided an overview of the NTCIR-19 Automatic Evaluation of LLMs 2 (AEOLLM-2) task, which aims to explore effective methods for automatically evaluating long-form responses generated by large language models (LLMs). AEOLLM-2 received a total of 91 runs from 10 teams, covering a diverse range of approaches to LLM evaluation. The final evaluation revealed several notable findings: (1) RSLIR achieved the best overall performance and obtained the highest scores on comprehensiveness, insight, and instruction following, while SCUNLP performed best on readability. (2) Spearman's rank correlation ($\rho$), Kendall's Tau ($\tau$), and accuracy generally showed consistent trends, although some differences in method rankings were observed, indicating that the three metrics capture complementary aspects of evaluation performance. (3) Among the four evaluation dimensions, instruction following and readability were relatively more challenging: instruction following showed the lowest best accuracy, while readability exhibited comparatively low rank correlations. These findings demonstrate that different evaluation dimensions pose distinct challenges and highlight the importance of using multiple metrics and fine-grained criteria when evaluating long-form responses. Looking ahead, we plan to further extend AEOLLM-2 with more diverse data and scenarios~\cite{chen2026benchmarking} to support more comprehensive and reliable evaluation of LLMs.


\bibliographystyle{ACM-Reference-Format}
\bibliography{sample-base}

\end{document}

%% file: intro.tex
\label{section: intro}


The Automatic Evaluation of LLMs 2 (AEOLLM-2) task is a core task in NTCIR-19 to support in-depth research on large language model (LLM) evaluation.

The rapid progress of LLMs such as GPT~\cite{achiam2023gpt}, ChatGLM~\cite{glm2024chatglm,zeng2022glm}, and Qwen~\cite{bai2023qwen} has attracted a lot of interest in Artificial General Intelligence within both academia and industry~\cite{chen2025enhancing,chen2026multi,chen2026simulating}. Alongside this advancement, a critical challenge has emerged: how to efficiently and effectively evaluate the performance of these models. Existing evaluation methods for LLMs can be broadly classified into two categories: manual evaluation and automatic evaluation. Manual evaluation involves engaging human annotators to directly assess the quality of responses generated by LLMs. Conversely, automatic evaluation relies on standard metrics and evaluation tools to assess model performance. In comparison to human evaluation, automatic evaluation mitigates the need for extensive human involvement, thereby reducing costs and enhancing the objectivity of the evaluation process \cite{chang2023survey}. As a result, investigating efficient and general automatic evaluation methods becomes increasingly significant and holds great promise.

However, existing automatic evaluation methods for LLMs still have the following limitations: (1) Limited task format. The majority of existing automatic LLM evaluation benchmarks only contain multiple-choice-format questions. Although this setting facilitates the computation of evaluation metrics, the multiple-choice format differs from real-world practical questions, which are usually open-ended without definite answers; (2) Limited evaluation criteria. Reference-based metrics~\cite{lin2004rouge, papineni2002bleu, zhang2019bertscore} (such as Rouge and BLEU) are widely used in many LLM tasks. These kinds of metrics cannot accurately reflect the quality of the results. Moreover, LLMs' strong memory capabilities enable them to rapidly learn and optimize any public benchmarks during training, which makes the reference-based metrics on these benchmarks useless for testing immediately. Based on these considerations, we proposed the Automatic Evaluation of LLMs (AEOLLM) task at NTCIR-18.

Through AEOLLM~\cite{chen2025overview}, we observed that most popular automatic evaluation approaches follow the LLMs-as-Judges paradigm~\cite{li2024llms, chen2026auto}, in which LLMs are used to assess the outputs generated by other LLMs. However, our findings revealed that these methods still remain limited in effectiveness when applied to long-form generation tasks, such as story generation. Meanwhile, there is a growing trend in the community toward using LLMs to generate \textbf{Deep Research Reports}~\cite{openai2025introducing}, which represents a new class of long-form generation tasks that meets real-world demand for structured and in-depth research content. These reports often exceed 1500 tokens  (8,000+ for scientific surveys) and require a high degree of factual accuracy, structured organization, and thematic coherence. Although a few recent studies have started to explore automatic evaluation techniques for such reports, the reliability of these methods has not been systematically validated. What's more, as summarized in Table~\ref{tab:my-table}, existing popular LLM-as-a-Judge benchmarks remain focused primarily on short to medium-length texts (typically less than 500 tokens), leaving long-form scenarios largely underexplored.

To address this gap, we propose a new subtask in AEOLLM-2: \textbf{Deep Research Evaluation}. In this subtask, participants will be provided with a set of deep research reports generated by LLMs. Their goal is to design automatic evaluation methods that can assess the quality of these reports. The results produced by each method will then be compared against human-annotated labels, allowing us to assess the reliability and effectiveness of automated evaluation techniques in the context of long-form text generation. We believe that this task will assist in improving the performance of LLM evaluation methods and advance the development of LLMs.

\begin{table}[t]
\caption{Statistics of candidate output lengths in existing LLM-as-a-judge benchmarks. \# Inst., Mean, and Med. denote instance count, mean length, and median length, respectively. Length is measured in tokens using the \texttt{cl100k\_base} tokenizer from \texttt{tiktoken}.}
\label{tab:my-table}
\resizebox{\columnwidth}{!}{%
\begin{tabular}{lrrr}
\hline
Benchmark                                 & \# Inst. & Mean  & Med. \\ \hline
MT-Bench~\cite{zheng2023judging}          & 4,796   & 194.0 & 142  \\
LLMBar~\cite{zeng2024evaluating}          & 419     & 106.0 & 49   \\
RewardBench~\cite{lambert2025rewardbench} & 8,108   & 194.8 & 115  \\
PreferenceBench~\cite{kim2024prometheus}  & 1,998   & 177.2 & 168  \\
RPR~\cite{pitis2024improving}             & 11,167  & 94.0  & 89   \\
JudgeBench~\cite{tan2025judgebench}       & 620     & 446.0 & 387  \\
RewardBench 2~\cite{malik2025rewardbench}  & 1,865   & 312.3 & 265  \\ \hline
\end{tabular}%
}
\end{table}

\begin{table}[]
\caption{NTCIR-19 AEOLLM-2 timeline.}
\label{timeline}
\resizebox{\columnwidth}{!}{%
\begin{tabular}{ll}
\hline
Dataset Release            & March 15, 2026              \\
Dry run Deadline           & July 15, 2026             \\
Formal run        & July 15, 2026-August 8, 2026 \\
Evaluation Results Release & August 8, 2026              \\ \hline
\end{tabular}
}
\end{table}

\begin{table}[]
\caption{NTCIR-19 AEOLLM-2 run statistics.}
\label{run}
\resizebox{\columnwidth}{!}{%
\begin{tabular}{llll}
\hline
Team     & Dry run & Formal run & Total \\ \hline
IITBHU   & 5       & 2          & 7     \\
ISLab    & 10      & 12         & 22    \\
LINKLab  & 6       & 2          & 8     \\
LLATMU   & 1       & 0          & 1     \\
RSLIR    & 12      & 11         & 23    \\
SCUNLP   & 12      & 1          & 13    \\
THUFAITH & 5       & 2          & 7     \\
TMUHLLA  & 3       & 1          & 4     \\
UBCS     & 3       & 0          & 3     \\
kucs17   & 1       & 2          & 3     \\ \hline
Total    & 58      & 33         & 91    \\ \hline
\end{tabular}
}
\end{table}

The timeline of the NTCIR-19 AEOLLM-2 task is shown in Table \ref{timeline}. This year, we received 91 runs from 10 teams in total. The statistics are given in Table \ref{run}. 

In summary, our contributions include: (1) Providing a practical dataset. (2) Offering an automated evaluation pipeline that facilitates performance testing of different methods. More details can be found at \url{https://huggingface.co/spaces/THUIR/AEOLLM}.

The structure of the remaining sections of this paper is described as follows. In Section 2, we present the overall design and evaluation framework of AEOLLM-2. In Section 3, we detail the implementation, including dataset construction and evaluation metrics. In Section 4, we present the participants' methods. In Section 5, we report the final evaluation results. Finally, we conclude the paper in Section 6.

%% file: frame.tex
\begin{figure}[t]
  \centering
  \includegraphics[width=\columnwidth]{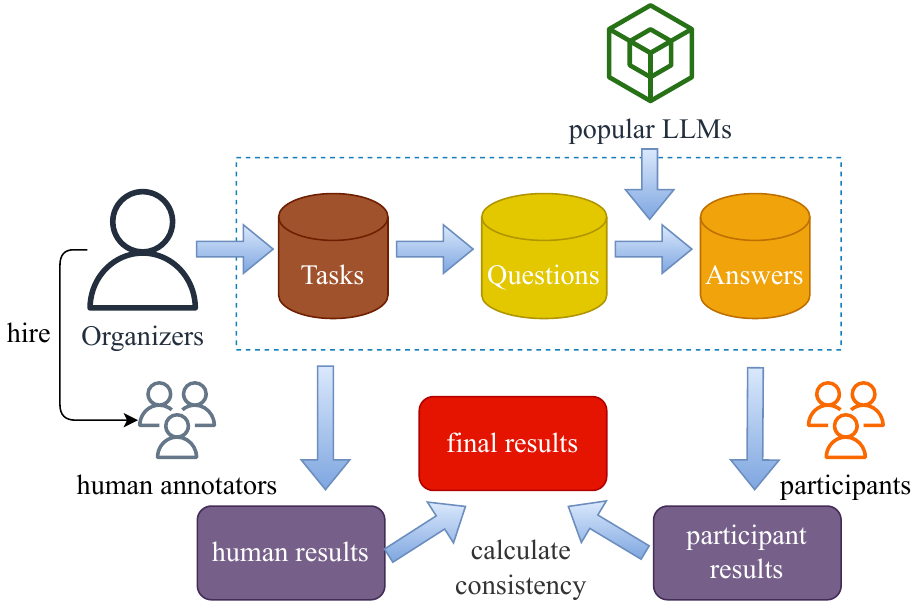}
  \caption{The overall framework of the AEOLLM-2 task.}
  \label{all_frame}
\end{figure}

Figure \ref{all_frame} illustrates the overall framework of the AEOLLM-2 task. As discussed in Section 1, AEOLLM-2 focuses on the evaluation of long-form text generation. We use DeepResearchBench~\cite{du2026deepresearch} as the data source for task construction.

Then, we choose a series of popular LLMs during the competition to generate answers. We manually annotate the answer sets for each question, which will be used as gold standards for evaluating the performance of different evaluation methods.

Finally, we collect evaluation results from participants on Answer sets for different Question sets under each task, and calculate consistency with manually annotated results. We use accuracy ($acc$), Kendall's tau ($\tau$) \cite{kendall1938new}, and Spearman correlation coefficient ($\rho$) \cite{lehman2013jmp}  as the evaluation metrics, which will be discussed next.

%% file: details.tex
\subsection{Datasets}

\subsubsection{Question and Prompt Construction}
We selected 40 questions in total, including 29 Chinese questions and 11 English questions. For each question, we designed 4 prompt variants that requested a deep research report on the same topic but were intentionally constructed to induce different levels and aspects of response quality.

Specifically, some prompts encouraged more detailed and informative reports, while others tended to produce simpler and less comprehensive responses. Similarly, some prompts emphasized balanced and neutral analysis, whereas others could introduce potential biases. This design allowed us to examine whether evaluation methods could reliably distinguish reports of different quality rather than relying primarily on surface-level fluency.

\subsubsection{Answer Generation}
For each prompt variant, we used four deep research systems, including Grok, Gemini, Perplexity, and Mita, to generate responses. As a result, each question produced 16 responses from four prompt variants and four systems, yielding a total of 640 question-answer instances. The median document length is
about 9,610 tokens.

\subsubsection{Human Annotation}

We evaluate each generated report along four dimensions: comprehensiveness, insight, instruction following, and readability. Comprehensiveness measures whether the report covers the key aspects and information required by the task. Insight evaluates the depth of analysis and the novelty of the presented viewpoints. Instruction following measures how accurately the report satisfies the explicit task requirements. Readability evaluates the overall clarity, fluency, and ease of understanding of the report.

To reduce annotation costs while maintaining reliable reference judgments, we adopt a human-in-the-loop annotation procedure for the relatively objective dimensions, including comprehensiveness, insight, and instruction following. For these dimensions, we first use retrieval-augmented generation (RAG) to retrieve relevant evidence and employ an LLM to produce preliminary scores. Human annotators then review the retrieved evidence and preliminary scores and revise the scores when necessary. This procedure reduces the amount of manual effort required while retaining human verification of the final annotations.

We treat readability as a subjective dimension and rely entirely on human annotation. For each task-document pair, we recruit two annotators with relevant backgrounds to independently evaluate the document on a 0--10 scale. The annotation considers three aspects: logical structure, presentation form, and bias checking. Logical structure evaluates the overall organization of the report, including section division, heading-content alignment, and reasoning coherence. Presentation form evaluates language fluency, conciseness, readability, and the coordination between textual and visual elements. Bias checking evaluates potential one-sidedness, logical fallacies, over-reliance on authority, unfair framing, and misleading statements. The three aspects are assigned document-specific weights according to the annotation guidelines, and their weighted sum is used as the final subjective score.

When the scores from the two annotators differ substantially according to a predefined adjudication rule, a third annotator is introduced to resolve the disagreement. We measure annotation agreement using the average-measure intraclass correlation coefficient (ICC), obtaining an $\mathrm{ICC}_{\mathrm{avg}}$ of 0.76, which indicates high agreement. The average score across the human annotations is used as the final reference judgment.

\subsubsection{Data Acquisition and Usage}
We divided the 640 instances into three parts:

1. train set: 25\% of the data was released as the training set, including annotations, for participants to reference when developing their evaluation methods.

2. dry run set: Another 25\% of the data was used for the Dry Run phase, which allowed participants to evaluate and refine their methods before the final evaluation.

3. formal run set: The remaining 50\% of the data was kept hidden and used for the Formal Run phase to determine the final evaluation results.

Both the training set and the test set can be downloaded from the provided link: \url{https://huggingface.co/datasets/THUIR/AEOLLM}.

\subsection{Evaluation Metrics}

We evaluate the agreement between participant scores and human annotations using three metrics: pairwise accuracy ($acc$), Spearman's rank correlation coefficient ($\rho$), and Kendall's rank correlation coefficient ($\tau$). These metrics are computed for the four evaluation dimensions, namely comprehensiveness, insight, instruction following, and readability, as well as for a weighted overall score.

For each question $q$ and answer $i$, the weighted overall score is computed using the weights specified for each evaluation dimension in the corresponding rubric:
\[
s_{q,i}^{\mathrm{overall}}
=
\frac{
\sum_{d \in \mathcal{D}} w_{q,d}s_{q,i,d}
}{
\sum_{d \in \mathcal{D}} w_{q,d}
},
\]
where $\mathcal{D}$ denotes the set of four evaluation dimensions, $s_{q,i,d}$ is the score of answer $i$ on dimension $d$, and $w_{q,d}$ is the corresponding weight defined in the rubric. The same weighting scheme is applied to both participant scores and human annotations.

\subsubsection{Accuracy ($acc$)}
Pairwise accuracy measures the agreement between the pairwise preferences derived from participant scores and human annotations. For each question, every pair of candidate answers is compared. Given two answers $i$ and $j$, their preference relation is defined as
\[
z_{i,j} = \operatorname{sign}(s_i-s_j),
\]
where $z_{i,j}\in\{-1,0,1\}$ indicates that answer $i$ is ranked below, tied with, or above answer $j$, respectively. The pairwise accuracy for question $q$ is computed as
\[
acc_q =
\frac{1}{|\mathcal{P}_q|}
\sum_{(i,j)\in\mathcal{P}_q}
\mathbb{I}
\left[
\hat{z}_{q,i,j}=z^{*}_{q,i,j}
\right],
\]
where $\mathcal{P}_q$ denotes the set of all valid answer pairs for question $q$, and $\hat{z}_{q,i,j}$ and $z^{*}_{q,i,j}$ denote the preference relations derived from participant scores and human annotations, respectively. A tied pair is considered correct only when both participant scores and human annotations indicate a tie.

\subsubsection{Spearman's Rank Correlation Coefficient ($\rho$)}
Spearman's rank correlation coefficient measures the agreement between the rankings derived from participant scores and human annotations. For each question, both sets of scores are converted into ranks, with tied scores assigned their average rank. Spearman's correlation is then computed as the Pearson correlation between the two rank vectors:
\[
\rho_q =
\frac{
\sum_i
(\hat{r}_{q,i}-\bar{\hat{r}}_q)
(r^{*}_{q,i}-\bar{r}^{*}_q)
}{
\sqrt{
\sum_i(\hat{r}_{q,i}-\bar{\hat{r}}_q)^2
}
\sqrt{
\sum_i(r^{*}_{q,i}-\bar{r}^{*}_q)^2
}
},
\]
where $\hat{r}_{q,i}$ and $r^{*}_{q,i}$ denote the ranks of answer $i$ derived from participant scores and human annotations, respectively.

\subsubsection{Kendall's Rank Correlation Coefficient ($\tau$)}
Kendall's rank correlation coefficient measures the consistency of pairwise ordering between participant scores and human annotations. Since tied scores may occur, we use Kendall's $\tau_b$, which accounts for ties:
\[
\tau_q =
\frac{C_q-D_q}
{
\sqrt{
(C_q+D_q+T_q)
(C_q+D_q+U_q)
}
},
\]
where $C_q$ and $D_q$ denote the numbers of concordant and discordant answer pairs, respectively. $T_q$ and $U_q$ denote the numbers of pairs tied only in the participant scores and only in the human annotations, respectively. Pairs tied in both sets of scores are not included in either $T_q$ or $U_q$.

For all three metrics, we first compute the metric independently for each question and then average the resulting values across all questions. Higher values indicate stronger agreement with human annotations.
\subsection{Answer Organization}

We follow a similar format as the ones used by most TREC~\cite{voorhees2005trec} submissions:

\begin{enumerate}
    \item White space is used to separate columns.
    \item The width of the columns in the format is not important, but it is important to have exactly seven columns per line with at least one space between the columns.
\end{enumerate}

These seven columns are \textbf{taskId questionId answerId comprehensiveness\_score insight\_score instruction\_following\_score readability\_score}.

\begin{itemize}

    \item The first column is the \textbf{taskId} (index different tasks).

    \item The second column is \textbf{questionId} (index different questions in the same task).

    \item The third column is \textbf{answerId} (index the answer provided by different LLMs to the same question).

    \item The fourth column is \textbf{comprehensiveness\_score} (index the comprehensiveness score).

    \item The fifth column is \textbf{insight\_score} (index the insight score).

    \item The sixth column is \textbf{instruction\_following\_score} (index the instruction\_following score).

    \item The seventh column is \textbf{readability\_score} (index the readability score).

\end{itemize}

%% file: method.tex
\begin{itemize}
    \item \textbf{RSLIR}~\cite{rslir2026} proposes OSODT, in which a DeepSeek-v4-Flash judge evaluates one dimension per call and aggregates criterion-level scores using the official rubric weights. The team found that changing the order of rubric criteria caused far larger score variation than repeated runs, so it averages the scores obtained under the original and the reversed criterion order.

    \item \textbf{ISLab}~\cite{islab2026} uses Gemma-4 31B as the judge and optimizes its scoring prompt automatically. The method combines TextGrad-style revision guided by scoring errors with ProCut-style pruning of low-contribution prompt segments. A revised prompt is accepted only if it reduces the loss on a held-out validation set, and pruning proves essential to prevent the performance loss caused by prompt bloat.

    \item \textbf{TMUHLLA}~\cite{tmuhlla2026} uses a local holistic Gemma-4 31B judge as the primary ranking signal and consults a secondary signal only to break exact ties. The secondary signal combines rubric-conditioned Qwen3-32B scores and text features with pairwise comparisons judged in both A/B orders. A comparison is kept only when both orders agree on the winner.

    \item \textbf{LINKLab}~\cite{linklab2026} is the only team to fine-tune a judge. It converts human scores into position-balanced A/B/tie preference labels and uses them to train a LoRA-adapted Qwen3-8B that predicts preferences on all four dimensions in a single generation. The judge compares all report pairs of each question using the full texts, and a Bradley--Terry model for each dimension aggregates the comparisons into final scores.

    \item \textbf{IITBHU}~\cite{iitbhu2026} adopts a lightweight zero-shot pipeline in which Gemini 3.1 Flash Lite scores every rubric sub-criterion of the full report in a single call, followed by weighted aggregation. Despite its simplicity, it achieves the second-best Kendall correlation on both insight and readability.

    \item \textbf{SCUNLP}~\cite{scunlp2026} scores each of the roughly twenty question-specific rubric criteria separately and combines them with the rubric weights, which yields continuous scores without ties. The team identifies ties as a major source of error for holistic judges. It also blends a regression on structural features into the readability score and corrects a length bias in instruction following, achieving the best readability results among all participants in the formal run.

    \item \textbf{THUFAITH}~\cite{thufaith2026} uses the averaged pointwise scores of two LLM judges as per-dimension anchors. For insight, four judges compare nearby reports in both presentation orders, and the pooled comparisons are fitted with a Davidson--Bradley--Terry model regularized toward the anchor.
\end{itemize}

%% file: result.tex
Table \ref{result} presents the performance comparison between the submitted methods and baseline models in the formal run on the reserved set. Based on the results, we make the following observations.

\begin{itemize}

\item \textbf{Comparing different methods,} RSLIR achieves the strongest overall performance, obtaining the best results on all three weighted overall metrics, with a Spearman correlation of 0.7071, a Kendall correlation of 0.5592, and an accuracy of 0.7796. RSLIR also consistently achieves the best performance on comprehensiveness, insight, and instruction following across all three metrics, demonstrating strong and balanced evaluation capability across these dimensions. In contrast, SCUNLP performs best on readability, achieving the highest Spearman correlation (0.5197), Kendall correlation (0.3827), and accuracy (0.6883). The second-best results are distributed across several methods, including ISLab, TMUHLLA, THUFAITH, IITBHU, and kucs17, indicating that different approaches exhibit different strengths across evaluation dimensions. Among the baseline methods, GLM-4.7 shows relatively competitive performance and achieves the second-best Spearman correlation on readability (0.4710). Nevertheless, the leading participant methods generally outperform the baselines, suggesting that specifically designed evaluation methods can provide clear improvements over directly using general-purpose LLMs as judges.

\item \textbf{Comparing different evaluation metrics,} the three metrics generally show consistent trends across methods and evaluation dimensions. For example, RSLIR achieves the best results under all three metrics for weighted overall, comprehensiveness, insight, and instruction following, while SCUNLP performs best across all three metrics for readability. The rankings of other methods are also largely consistent across Spearman's rank correlation ($\rho$), Kendall's Tau ($\tau$), and accuracy. However, some differences can still be observed. For instance, on comprehensiveness, ISLab achieves the second-best $\rho$ and $\tau$, while THUFAITH obtains the second-best accuracy. These differences indicate that although the three metrics are generally consistent, they capture slightly different aspects of evaluation performance. This suggests that considering multiple metrics provides a more reliable and comprehensive assessment of automatic evaluation methods.

\item \textbf{Comparing different evaluation dimensions,} instruction following and readability appear to be relatively more challenging than comprehensiveness and insight. For instruction following, the best accuracy reaches only 0.6350, substantially lower than the best accuracy for comprehensiveness (0.7533) and insight (0.7425). Readability, meanwhile, shows particularly low rank correlations, with the best Spearman and Kendall correlations reaching only 0.5197 and 0.3827, respectively. In comparison, the best correlation scores for comprehensiveness and insight are considerably higher. These results suggest that instruction following and readability present different challenges to automatic evaluators: accurately judging instruction adherence remains difficult, while consistently ranking responses by readability is also challenging. This further highlights the need for fine-grained evaluation across multiple dimensions when assessing long-form responses.

\end{itemize}